\documentclass[12pt, a4paper]{amsart}
\usepackage{amsmath}
\usepackage{amsxtra}
\usepackage{amscd}
\usepackage{amsthm}
\usepackage{amsfonts}
\usepackage{amssymb}
\usepackage {pstricks}
\usepackage{pstricks,pst-node}

\theoremstyle{definition}

\theoremstyle{remark}

\numberwithin{equation}{section}

\newcommand{\R}{{\mathbb R}}

\newcommand{\barh}{{\bar h}}
\newcommand{\cA}{{\mathcal A}}

\newcommand{\cM}{{\mathcal M}}
\newcommand{\cR}{{\mathcal R}}

\newcommand{\bM}{{\mathbf M}}

\newcommand{\lAm}{{_l\cA^m}}
\newcommand{\rAm}{{\cA_r^m}}

\newcommand{\g}{{\mathbf g}}

\begin{document}


\title[Exact solutions of noncommutative Einstein equations]
{Exact solutions of noncommutative vacuum Einstein field equations and plane-fronted gravitational waves}

\author{Ding Wang}
\address{Institute of Mathematics,
Academy of Mathematics and Systems Science, Chinese Academy of
Sciences, Beijing 100190, China}
\email{wangding@amss.ac.cn}

\author{R. B. Zhang}
\address{School of Mathematics
and Statistics, University of Sydney, Sydney, NSW 2006, Australia}
\email{rzhang@maths.usyd.edu.au}

\author{Xiao Zhang}
\address{Institute of Mathematics,
Academy of Mathematics and Systems Science, Chinese Academy of
Sciences, Beijing 100190, China}
\email{xzhang@amss.ac.cn}

\begin{abstract}
We construct a class of exact solutions of the noncommutative
Einstein field equations in the vacuum, which are noncommutative
analogues of the plane-fronted gravitational waves in classical
gravity.
\end{abstract}
\maketitle

%
%
\section{Introduction}

There have been intensive research activities on noncommutative
relativity in recent years.  In particular, several tentative
proposals \cite{R6, R7, AMV, BMS, R4, CTZZ, ZZ} for a theory of
noncommutative relativity   have been put forward. In \cite{R7}
noncommutativity was introduced into gravity by deforming the
diffeomorphism algebra. In \cite{R6, R4} general relativity on a
noncommutative spacetime is treated as a noncommutative gauge
theory. Very recently, the authors of \cite{BM} explored a possible
moving frame formalism for a noncommutative geometry on the Moyal
space as the first step toward setting up a framework for
noncommutative general relativity. Much work has also been done to
investigate implications of spacetime noncommutativity to black hole
physics \cite{WZZ, R1, R4}.

In the papers \cite{CTZZ, ZZ}, a formalism for spacetime
quantisation was proposed, which made use of isometric embeddings
\cite{N, Cl, Gr} of spacetime into pseudo-Euclidean spaces. In this
formalism, one first finds a global embedding of a spacetime into
some pseudo-Euclidean space, whose existence is guaranteed by
theorems of Nash, Clarke and Greene \cite{N, Cl, Gr}. Then one
quantises the spacetime following the strategy of deformation
quantisation \cite{BFFLS, Ko} by deforming \cite{Ge} the algebra of
functions in the pseudo-Euclidean space to a noncommutative
associate algebra known as the Moyal algebra. Through this
mechanism, classical spacetime metrics will deform to ``quantum''
noncommutative metrics which acquire quantum fluctuations. In
particular, certain anti-symmetric components arise in the deformed
metrics, which involve the Planck constant and vanish in the
classical limit.

The theory of \cite{CTZZ, ZZ, WZZ} can be formulated in an intrinsic
way, free of the use of emebeddings. This theory retains the notions
of connections and curvatures in the noncommutative setting in a
mathematically consistent manner. In particular, the quantum
deformed noncommutative Ricci curvatures can be defined in a unique
way. This enabled one to develop a noncommutative analogue of the
Einstein field equations \cite{CTZZ, ZZ}.

It is important to solve the noncommutative Einstein field equations
to construct quantum noncommutative spacetimes. In \cite{WZZ}, we
obtained noncommutative analogues of Schwarzschild spacetime and
de-Sitter Schwarzschild spacetime, which are approximate solutions
of the noncommutative Einstein field equations exact to the first
order of the deformation parameter. Quantum corrections to the area
law of black hole entropy was observed for these solutions.

In this letter we construct a class of {\em exact} solutions of the
noncommutative Einstein field equations in the vacuum. These
solutions are quantum deformations of the plane-fronted
gravitational waves first constructed by Brinkmann in 1925 \cite{Br}
and have since been rediscovered several times (e.g. \cite{ER, Ro,
EK}). Our solutions are noncommutative gravitational analogues of
electromagnetic plane waves. We expect them to have an important
role to play in future investigations of quantum gravity.

Fuzzy pp-waves were constructed by Madore, Maceda and Robinson in
\cite{MMR}. These authors start with a given classical solution of
the Einstein field equation in the vacuum and construct a
noncommutative algebra and a differential calculus which supported
the metric. The corresponding noncommutative scalar curvature was
however nonzero. In general the quantum deformed metrics of most
classical spacetimes satisfy the field equations only up to some
order of the Planck constant.

Exact (that is, not approximate) solutions of noncommutative
Einstein field equations do not seem to have been investigated much
in the literature. Presumably this is partly due to the fact that
many of the proposals of noncommutative relativity are based on
intuition. Much work has been done to investigate corrections to
physically relevant quantities to the first order in the deformation
parameter within the frameworks of the various proposals. However,
to go beyond the first order approximation, one will need a
mathematically more rigorous theory. In particular, we can only
investigate exact solutions when precisely formulated noncommutative
Einstein field equations are given.

We mention that even within the mathematically rigorous formulations
like those of \cite{MMR} and \cite{CTZZ, ZZ}, the mathematical
complexities introduced by spacetime noncommutativity makes it
extremely difficult to study exact solutions of the noncommutative
Einstein field equations. Thus it is quite remarkable that the
quantum deformed plane-fronted gravitational waves constructed here
solve the noncommutative vacuum Einstein field equations exactly.

\section{Noncommutative Einstein equations}

In order to set up the noncommutative Einstein equations, we need to
have a noncommutative differential geometry which retains the
notions of metric, connection and curvature. Such a theory was
constructed in \cite{CTZZ, ZZ}. We describe the theory very briefly
here; details can be found in \cite{CTZZ, ZZ}.

\subsection{A local noncommutative differential geometry}\label{sect-embedding}
Let $U$ be a domain in $\R^n$ with natural coordinates $\{x^0,
\cdots, x^n\}$. Let $\barh$ be a real indeterminate, and denote by
$\R[[\barh]]$ the ring of formal power series in $\barh$. Let $\cA$
be the set of formal power series in $\barh$ with coefficients being
real smooth functions on $U$. Namely, every element of $\cA$ is of
the form $\sum _{i\ge 0} f_i\barh^i$ where $f_i$ are smooth
functions on $U$. Then $\cA$ is an $\R[[\barh]]$-module.

Given any two smooth functions $u$ and $v$ on $U$, we denote by $u
v$ the usual point-wise product of the two functions. We also define
their star-product (or more precisely, Moyal product) $u\ast v$ on
$U$ by
\begin{eqnarray}\label{multiplication}
(u\ast v)(x) = \lim _{x'\rightarrow x}\ \exp{\left(\barh \sum _{i j}
\theta _{i j}\partial _i\partial _j^\prime\right)}u(x) v(x'),
\end{eqnarray}
where $\partial _i=\frac{\partial}{\partial x^i}$,  and $(\theta _{i
j})$ is a constant skew symmetric $n\times n$ matrix. It is well
known that such a multiplication is associative. Thus $\cA$ equipped
with the Moyal product is a deformation of the algebra of functions
on $U$ in the sense of \cite{Ge}. Since $\theta$ is constant, the
Leibniz rule remains valid in the present case:
\[ \partial _i(u\ast v)= \partial _i u\ast v + u\ast \partial _i v. \]

In noncommutative geometry \cite{Con}, the associative algebra $\cA$
is regarded as defining some {\em quantum deformation of the region
$U$}, and finitely generated projective modules over $\cA$ are
regarded as (spaces of sections of) noncommutative vector bundles on
the quantum deformation of $U$ (defined by the noncommutative
algebra $\cA$). Given an integer $m>n$, we let $\lAm$ (resp. $\rAm$)
be the set of $m$-tuples with entries in $\cA$ written as rows
(resp. columns). We shall regard $\lAm$ (resp. $\rAm$) as a left
(resp. right) $\cA$-module with the action defined by multiplication
from the left (resp. right). More explicitly, for
$v=\begin{pmatrix}a_1 & a_2 & \dots & a_m\end{pmatrix}\in\lAm$, and
$b\in\cA$, we have $b \ast v =
\begin{pmatrix}b\ast a_1 & b\ast a_2 & \dots & b\ast a_m\end{pmatrix}$.
Similarly for $w=\begin{pmatrix}a_1 \\  a_2 \\ \vdots \\
a_m\end{pmatrix}\in\rAm$, we have $w\ast b = \begin{pmatrix}a_1\ast b \\
a_2\ast b\\ \vdots \\ a_m\ast b\end{pmatrix}$. Let $\bM_m(\cA)$ be
the set of $m\times m$-matrices with entries in $\cA$. We define
matrix multiplication in the usual way but by using the Moyal
product for products of matrix entries, and still denote the
corresponding matrix multiplication by $\ast$. Now for $A=(a_{i j})$
and $B=(b_{i j})$, we have $(A\ast B)=(c_{i j})$ with $c_{i j} =
\sum_{k} a_{i k}\ast b_{k j}$. Then $\bM_m(\cA)$ is an
$\R[[\barh]]$-algebra, which has a natural left (resp. right) action
on $\rAm$ (resp. $\lAm$).

A finitely generated projective left (reps. right) $\cA$-module is
isomorphic to some direct summand of $\lAm$ (resp. $\rAm$) for some
$m<\infty$. If $e\in \bM_m(\cA)$ satisfies the condition $e\ast
e=e$, that is, it is an idempotent, then
\[
\cM=\lAm\ast e := \{v\ast e \mid v\in \lAm \}, \quad
\tilde\cM=e\ast\rAm := \{e\ast w \mid \in \rAm \}
\]
are respectively projective left and right $\cA$-modules.
Furthermore, every projective left (right) $\cA$-module is
isomorphic to an $\cM$ (resp. $\tilde\cM$) constructed this way by
using some idempotent $e$.

As the noncommutative geometries on the left module $\cM$ and right
module $\tilde \cM$ are equivalent, we need only to investigate
$\cM$. Let \[\omega_i = -
\partial_i e\] be the canonical connections on $\cM$. The covariant
derivative on the noncommutative bundle $\cM$ is given by
\[
\nabla_i \zeta = \partial_i\zeta + \zeta \ast\omega_i , \quad
\forall \zeta \in \cM.
\]
The {\em curvature} of $\cM$ associated with the connection $\omega_i$
is given by
\[
\cR_{i j}=\partial_i \omega_j - \partial_j \omega_i -
[\omega_i, \ \omega_j]_\ast,
\]
where $[\omega_i, \omega_j]_\ast = \omega_i\ast \omega_j
-\omega_j\ast \omega_i$ is the commutator.

Let $\eta=diag(1, \dots, 1, -1, \dots, -1)$ be a diagonal $m\times
m$ matrix with $p$ of the diagonal entries being $1$, and $q=m-p$ of
them being $1$ for some $p$.  The {\em fibre metric} is the
$\cA$-bilinear map
\begin{eqnarray}
\g: \cM\otimes_{\R[[\barh]]}\tilde\cM \longrightarrow \cA, \quad
v\otimes w \mapsto v * w,
\end{eqnarray}
where for any
$v=\begin{pmatrix} v_1 & \dots & v_m \end{pmatrix}\in\cM$
and $w=\begin{pmatrix} w_1 \\
\vdots \\ w_m \end{pmatrix} \in\tilde\cM$, we have $v*
w=\sum_{i=1}^m v_i* w_i$. The metric compatibility of the connection
and also the Bianchi identities for the Riemannian curvature were
discussed in \cite{CTZZ}.

In certain situations, we may regard $\cM$ and $\tilde\cM$ as
noncommutative tangent bundles of some noncommutative space.
Consider the case when there exists a finite set of $\cA$-generators
$E_i$ ($i=1, 2, \dots, n$) of $\cM$ with the following properties.
The column vectors $\eta (E_i)^t$ (where $(E_i)^t$ are the
transposes of $E_i$) generate $\tilde\cM$, and the $n\times n$
matrix $(\g_{i j})$ with
\begin{eqnarray}\label{fibre-metric}
\g_{i j}= \g(E_i, \eta(E_j)^t)
\end{eqnarray}
is invertible over $U$.   Then the idempotent $e$ is given by
$e=\eta (E_i)^t  \ast \g^{i j} \ast E_j$. In this case, we call the
matrix $\g:=(\g_{i j})$ the {\em metric}.

We may consider the components of the curvature $\cR_{i j}$:
\[
R^l _{k i j}=E_k \ast \cR_{i j} \ast \tilde E ^l, \quad
R _{l k i j}=R^p _{k i j} \ast \g _{pl},
\]
where
$
\tilde E^i=\eta(E_j)^t\ast \g ^{j i}, \quad
E^i=\g ^{i j}\ast E_j
$, for $i=1, 2, \dots, n$.

A new feature is that there are two consistent ways to contract
$\cR_{i j}$, leading to two distinct noncommutative Ricci curvatures
$R^i _j$ and $\Theta ^i _j$ respectively defined by
\begin{eqnarray*}
R^i _j = E^i \ast \cR_{p j} \ast \tilde E ^p, \quad
\Theta ^i _j = E^p \ast \cR_{j p} \ast \tilde E ^i.
\end{eqnarray*}
Correspondingly there are two scalar curvatures $R=R^i_i$
and $\Theta=\Theta^i_i$. Both $R^i _j$ and $\Theta ^i _j$
reduce to the usual Ricci curvature
in the commutative limit.

We now state the {\em noncommutative Einstein field equations in the
vacuum} (for unknowns $E_i$) in this theory, which are given by
\begin{eqnarray}\label{vacuum}
R^i _j = 0, \quad  \Theta ^i _j = 0, \quad \text{for all $i, j$}.
\end{eqnarray}
The aim of this note is to construct exact solutions of the
equations.

\subsection{Embedded noncommutative spaces}
Embedded noncommutative spaces are elementary and manifestly
consistent realisations of the local differential geometry discussed
above. Given $X=\begin{pmatrix} X^1 & X^2 &\dots & X^m
\end{pmatrix}$ in $\lAm$, we define an
$n\times n$ matrix $(\g_{ij})_{i, j=1, 2, \dots, n}$ with entries
\begin{eqnarray}\label{metric}
\g_{i j}=\sum_{\alpha=1}^m \partial_i X^\alpha  \ast \eta_{\alpha
\beta}  \ast \partial_j X^\beta,
\end{eqnarray}
where $\eta_{\alpha \beta}=\pm \delta_{\alpha \beta}$ are the matrix
elements of the diagonal matrix $\eta$.

The matrix $\g=(\g_{i j})$ is invertible over $U$ if and only if
$\g|_{\barh=0}$ is invertible. We denote the inverse matrix of $\g$
by $(\g^{i j})$. In this case, $X$ reduces to an embedded space with
metric $\g|_{\barh=0}$ in the commutative limit with $\theta=0$.
Therefore, we call $X$ a {\em noncommutative space} embedded in
$\cA^m$ in analogy to the classical case.

Let $ E_i=\partial_i X$ for $i=1, 2, \dots, n$. Then
$(E_i)^t= \begin{pmatrix}\partial_i X^1 \\
\partial_i X^2
\\ \vdots \\ \partial_i X^m\end{pmatrix}$.
The matrix
\[
e=  \eta (E_i)^t  \ast \g^{i j} \ast  E_j
\]
is an idempotent and satisfies $E_i \ast  e = E_i$ and
$e*\eta(E_i)^t = \eta(E_i)^t$ for all $i$. The left (resp. right) projective
$\cA$-module $\cM={_l}\cA^m*e$ (resp.
$\cM=e*\cA_r^m$) associated to $e$ is the quantised
left (resp. right) tangent bundle of the embedded noncommutative space.
It is easy to show that the metric defined by the $\cA$-bilinear map
\eqref{fibre-metric} agrees with \eqref{metric}  in the present
case.

We may cast the formulation of the embedded noncommutative space
into a more familiar form. The connection is now given by
\[
\nabla _i E_j =\Gamma _{ij} ^k \ast E _k,
\]
where $\Gamma _{ij}^k$ can be explicitly described in the following
way. Let $\Gamma_{i j l} =\Gamma_{i j } ^k \ast \g_{kl}$. We have
\[
 \Gamma_{i j l} =
{}_c\Gamma_{i j l} + \Upsilon _{i l j}+\Upsilon _{j i l}-\Upsilon
_{l j i},
\]
with
\begin{eqnarray*}
_c\Gamma_{i j l} = \frac{1}{2}\left(\partial_i \g_{j l} +
\partial_j \g_{l i}
-\partial_l \g_{j i} \right),
\quad\Upsilon_{i j l} = \frac{1}{2}
\left(\partial_i E_j \ast  \eta(E_l)^t - E_l \ast \eta \partial_i
(E_j)^t\right),
\end{eqnarray*}
where the object $\Upsilon_{i j l}$ is called the noncommutative torsion
in \cite{CTZZ}. The curvatures are given by
\begin{eqnarray*}
&&R_{k i j}^l = -\partial_j\Gamma_{i k}^l - \Gamma_{i k}^p  \ast
\Gamma_{j p}^l + \partial_i\Gamma_{j k}^l +\Gamma_{j k}^p  \ast
\Gamma_{i p}^l,\\
&&R^i_j= \g^{i k}\ast R^p_{k p j}, \quad \Theta^l_p = \g^{i k}\ast
R^l_{k p i}.
\end{eqnarray*}
It was shown in \cite{CTZZ, ZZ} that
the two noncommutative scalar curvatures $R$ and $\Theta$ coincide
in the present case.

\section{Exact solutions of noncommutative Einstein field equations}

We shall now construct a class of exact solutions of the
noncommutative vacuum Einstein field equations. The solutions are
quantum deformed analogues of plane-fronted gravitational waves
\cite{Br, ER, Ro, EK}.

Let $(\theta _{ij})$ be an arbitrary constant skew symmetric $4
\times 4$ matrix, and endow the space of functions of the variables
$(x, y, u , v)$ with the Moyal product defined with respect to
$(\theta _{ij})$. We denote the resulting noncommutative algebra by
$\cA$.

Now we consider a noncommutative space $X$ embedded in $\cA^6$ by a
map of the form
\begin{eqnarray}\label{embedding}
X =\left(x, y, \frac{Hu+u+v}{\sqrt{2}},
\frac{H-\frac{u^2}{2}}{\sqrt{2}}, \frac{H u-u+v}{\sqrt{2}},
\frac{H+\frac{u^2}{2}}{\sqrt{2}}\right),
\end{eqnarray}
where, needless to say, the component functions are elements of
$\cA$.

Let us take $\eta=diag(1, 1, 1, 1, -1, -1)$, and construct the
noncommutative metric $\g$ by using the formula \eqref{metric} for
this embedded noncommutative space. A very lengthy calculation
yields the following result:
\begin{eqnarray*}
\begin{aligned}
\g=&
\begin{pmatrix}
 1 & 0 & -\barh (\theta _{yu} H_{xy}+\theta _{xu} H_{xx}) & 0 \\
 0 & 1 & -\barh (\theta _{yu} H_{yy}+\theta _{xu} H_{xy}) & 0 \\
 \barh (\theta _{yu} H_{xy}+\theta _{xu} H_{xx}) &
 \barh (\theta _{yu} H_{yy}+\theta _{xu}
   H_{xy}) & 2 H & 1 \\
 0 & 0 & 1 & 0
\end{pmatrix}.
\end{aligned}
\end{eqnarray*}
It is useful to note that in the classical limit with all $\theta_{i
j}=0$, the metric has Minkowski signature. In fact it reduces to the
matrix $\begin{pmatrix}
1 & 0 & 0 & 0\\
0 & 1 & 0 & 0\\
0 & 0 & 2H & 1\\
0 & 0 & 1 & 0
\end{pmatrix}$, which diagonalises to $diag(1, 1, H+\sqrt{1+H^2},
H-\sqrt{1+H^2})$. Further tedious computations produce the following
inverse metric:
\begin{eqnarray*}
\begin{aligned}
\g^{-1}=&
\begin{pmatrix}
 1 & 0 & 0 & \barh (\theta _{yu} H_{xy}+\theta _{xu} H_{xx}) \\
 0 & 1 & 0 & \barh (\theta _{yu} H_{yy}+\theta _{xu} H_{xy}) \\
 0 & 0 & 0 & 1\\
 -\barh (\theta _{yu} H_{xy}+\theta _{xu} H_{xx}) &
 -\barh (\theta _{yu} H_{yy}+\theta _{xu}
   H_{xy}) & 1 & g^{44}
\end{pmatrix}
\end{aligned}
\end{eqnarray*}
with
\begin{eqnarray*}
\begin{aligned}
g^{44}=&-\barh^2 \left(\theta _{yu} H_{yy}+\theta _{xu}
H_{xy}\right)\ast \left(\theta _{yu} H_{yy}+\theta _{xu}
H_{xy}\right)\\
&-\barh^2 \left(\theta _{yu} H_{xy}+\theta
_{xu}H_{xx}\right)\ast\left(\theta _{yu} H_{xy}+\theta
_{xu}H_{xx}\right)-2 H.
\end{aligned}
\end{eqnarray*}
Using these formulae we can compute $\Gamma _{ijk}$ and $\Gamma^k
_{ij}$, the nonzero components of which are given below:
\begin{eqnarray*}
\begin{aligned}
\Gamma_{113}=&-\barh(\theta _{yu} H_{xxy}+\theta _{xu}H_{xxx}),\\
\Gamma_{123}=&\Gamma_{213}= -\barh (\theta _{yu} H_{xyy}+\theta _{xu} H_{xxy}),\\
\Gamma_{133}=&\Gamma_{313}= H_x-\barh (\theta _{yu} H_{xyu}+\theta _{xu} H_{xxu}),\\
\Gamma_{223}=&-\barh (\theta _{yu} H_{yyy}+\theta _{xu} H_{xyy}),\\
\Gamma_{233}=&\Gamma_{323}=H_y-\barh (\theta _{yu} H_{yyu}+\theta _{xu} H_{xyu})\\
\Gamma_{331}=&-H_x, \quad \Gamma_{332}=-H_y,\\
\Gamma_{333}=&H_u-\barh(\theta _{yu} H_{yuu}+\theta_{xu} H_{xuu});\\
\Gamma_{11}^4=&-\barh (\theta _{yu} H_{xxy}+\theta _{xu}H_{xxx}),\\
\Gamma_{12}^4=&\Gamma_{21}^4= -\barh (\theta _{yu} H_{xyy}+\theta _{xu} H_{xxy})\\
\Gamma_{13}^4=&\Gamma_{31}^4=H_x-\barh (\theta _{yu} H_{xyu}+\theta _{xu} H_{xxu})\\
\Gamma_{22}^4=&-\barh(\theta _{yu} H_{yyy}+\theta _{xu} H_{xyy}),\\
\Gamma_{23}^4=&\Gamma_{32}^4=H_y-\barh (\theta _{yu} H_{yyu}+\theta _{xu} H_{xyu}),\\
\Gamma_{33}^1=&-H_x,\quad
\Gamma_{33}^2=-H_y,\\
\Gamma_{33}^4=&-H_x \ast \barh (\theta _{yu} H_{xy}+\theta_{xu} H_{xx})
-H_y \ast\barh (\theta _{yu}H_{yy}+\theta _{xu} \ H_{xy})\\
&+H_u-\barh (\theta _{yu} H_{yuu}+\theta _{xu}H_{xuu}).
\end{aligned}
\end{eqnarray*}

Remarkably, explicit formulae for curvatures can also be obtained,
even though the noncommutativity of the $\ast$-product complicates
the computations enormously. As an example, we give the computation
of $R_{3313}$ here:
\begin{eqnarray*}
\begin{aligned}
R_{3313}=&\frac{\partial \Gamma_{33}^p}{\partial
x}\ast\g_{p3}-\frac{\partial \Gamma_{13}^p}{\partial
u}\ast\g_{p3}+\Gamma_{33}^p\ast\Gamma_{1p3}-\Gamma_{13}^p\ast\Gamma_{3p3}\\
=&\frac{\partial \Gamma_{33}^1}{\partial
x}\ast\g_{13}+\frac{\partial \Gamma_{33}^2}{\partial
x}\ast\g_{23}+\frac{\partial \Gamma_{33}^4}{\partial
x}\ast\g_{43}-\frac{\partial \Gamma_{13}^4}{\partial
u}\ast\g_{43}\\
&+\Gamma_{33}^1\ast\Gamma_{113}+\Gamma_{33}^2\ast\Gamma_{123}\\
=& -H_{xx}\ast(-\barh \left(\theta -{yu} H_{xy}+\theta _{xu}
H_{xx}\right))\\
&-H_{xy}\ast(-\barh \left(\theta _{yu} H_{yy}+\theta _{xu}
H_{xy}\right))\\
&+\frac{\partial \Gamma_{33}^4}{\partial x}- H_{xu}+\barh \left(\theta
_{yu} H_{xyuu}+\theta _{xu}
H_{xxuu}\right)\\
&+(-H_x)\ast( -\barh \left(\theta _{yu} H_{xxy}+\theta _{xu}
H_{xxx}\right))\\
&+(-H_y)\ast( -\barh \left(\theta _{yu} H_{xyy}+\theta -{xu}
H_{xxy}\right))=0.
\end{aligned}
\end{eqnarray*}
The other components of the curvature can be obtained in the same
way. We have
\begin{eqnarray*}
\begin{array}{lll}
&R_{1313}=-R_{1331}=-H_{xx}, & R_{1323}=-R_{1332}=-H_{xy},\\
&R_{2313}=-R_{2331}=-H_{xy}, & R_{2323}=-R_{2332}=-H_{yy},\\
&R_{3113}=-R_{3131}=H_{xx},  & R_{3123}=-R_{3132}=H_{xy},\\
&R_{3213}=-R_{3231}=H_{xy},  & R_{3223}=-R_{3232}=H_{yy},\\
&R_{3331}=-R_{3313}=0,       & R_{3332}=-R_{3323}=0.
\end{array}
\end{eqnarray*}

Thus the nonzero components of $R^l _{ijk}$ are
\begin{eqnarray*}
\begin{array}{lll}
&R_{113}^4=-R_{131}^4=H_{xx}, \;\;\;\;\;\;\;\;R_{123}^4=-R_{132}^4=H_{xy},\\
&R_{213}^4=-R_{231}^4=H_{xy}, \;\;\;\;\;\;\;\;R_{223}^4=-R_{232}^4=H_{yy},\\
&R_{313}^1=-R_{331}^1=-H_{xx},\;\;\;\;\;R_{313}^2=-R_{331}^2=-H_{xy},\\
&R_{323}^1=-R_{332}^1=-H_{xy},\;\;\;\;\;R_{323}^2=-R_{332}^2=-H_{yy},\\
&R_{313}^4=-R_{331}^4=-H_{xx}\ast\barh (\theta _{yu}H_{xy}+\theta _{xu} H_{xx})\\
&\;\;\;\;\;\;\;\;\;\;\;\;\;\;\;\;\;\;\;\;\;\;\;\;\;\;
-H_{xy}\ast\barh (\theta_{yu} H_{yy}+\theta _{xu} H_{xy}),\\
&R_{323}^4=-R_{332}^4=-H_{xy}\ast\barh (\theta _{yu}H_{xy}+\theta _{xu} H_{xx})\\
&\;\;\;\;\;\;\;\;\;\;\;\;\;\;\;\;\;\;\;\;\;\;\;\;\;\;
-H_{yy}\ast\barh (\theta _{yu} H_{yy}+\theta _{xu}H_{xy}).
\end{array}
\end{eqnarray*}
From these formulae, we obtain the nonzero components of the Ricci
curvature:
\begin{eqnarray}\label{Equation}
R_3^4=\Theta_3^4=-H_{xx}-H_{yy}.
\end{eqnarray}

Thus the noncommutative vacuum Einstein field equations
(\ref{vacuum}) are satisfied if and only if the following equation
holds:
\begin{eqnarray}
H _{xx} +H _{yy} =0. \label{vacuum-pp}
\end{eqnarray}

Solutions of this linear equation for $H$ exist in abundance. Each
solution leads to an exact solution of the noncommutative vacuum
Einstein field equations. If we set $\theta$ to zero, we recover
from such a solution the plane-fronted gravitational wave \cite{Br,
ER, Ro, EK} in classical general relativity. Thus we shall call such
a solution of (\ref{vacuum}) a {\em plane-fronted noncommutative
gravitational wave}.

It is clear from \eqref{Equation} that plane-fronted noncommutative
gravitational waves satisfy the additivity property. Explicitly, if
the noncommutative metrics of
\begin{eqnarray*}
X^{(i)} =\left(x, y, \frac{H_iu+u+v}{\sqrt{2}},
\frac{H_i-\frac{u^2}{2}}{\sqrt{2}}, \frac{H_iu-u+v}{\sqrt{2}},
\frac{H_i+\frac{u^2}{2}}{\sqrt{2}}\right), \quad i=1, 2,
\end{eqnarray*}
are plane-fronted noncommutative gravitational waves, we let
$H=H_1+H_2$, and set
\begin{eqnarray*}
X=\left(x, y, \frac{Hu+u+v}{\sqrt{2}},
\frac{H-\frac{u^2}{2}}{\sqrt{2}}, \frac{H u-u+v}{\sqrt{2}},
\frac{H+\frac{u^2}{2}}{\sqrt{2}}\right).
\end{eqnarray*}
Then the noncommutative metric of $X$ is also a plane-fronted
noncommutative gravitational wave. This is a rather nontrivial fact
since the noncommutative Einstein field equations are highly
nonlinear in $\g$, and it is extremely rare to have this additivity
property.

At this point, it is appropriate to point out that the embedding
\eqref{embedding} is only used as a device for constructing the
metric and the connection, from which the curvatures are derived.
However, we should observe the power of embeddings in solving the
noncommutative Einstein field equations. Without using the embedding
\eqref{embedding}, it would be very difficult to come up with
elegant solutions like what we have obtained here.

\section{Discussions}

Working within the framework of the noncommutative Riemannian
geometry of \cite{CTZZ}, we have obtained in this paper exact
solutions of the quantum noncommutative vacuum Einstein field
equations, which are noncommutative analogues of the plane-fronted
gravitational waves in classical general relativity \cite{Br, ER,
Ro, EK}.

In the classical setting, the plane-fronted gravitational waves
model spacetimes moving at the speed of light and radiating energy.
Furthermore,  Penrose \cite{Pe} observed that near a null geodesic,
every spacetime can be blown up so that the given null geodesic
becomes the covariantly constant null geodesic congruence of a plane
wave. We expect the plane-fronted noncommutative gravitational waves
to play a similar role. It will be very interesting to investigate
the physical applications of these solutions.

It is quite striking that the quantum noncommutative Einstein field
equations \cite{CTZZ}, complicated as they are, admit explicit exact
solutions as simple as the ones constructed here. This indicates the
promise of the theory of noncommutative Riemannian geometry proposed
in \cite{CTZZ}. We hope that the theory will develop into a coherent
framework for studying the structure of spacetime at the Planck
scale.

\bigskip

\noindent{\bf Acknowledgement}: X. Zhang wishes to thank the School
of Mathematics and Statistics,
University of Sydney for the hospitality during his visits when part
of this work was carried out. Partial financial support from the
Australian Research Council, National Science Foundation of China
(grants 10421001, 10725105, 10731080), NKBRPC (2006CB805905) and the
Chinese Academy of Sciences is gratefully acknowledged.

\end{document}